\documentclass[10pt,aps,prl,amsmath,amssymb,twocolumn,letterpaper,%
         superscriptaddress,footinbib,showpacs,%
         citeautoscript,floatfix]{revtex4-1}
\usepackage[usenames,dvipsnames]{color}
\usepackage{graphicx}
\usepackage{microtype}
\usepackage{ifthen}
\usepackage[bookmarks=false,colorlinks]{hyperref}
\hypersetup{
    linkcolor=magenta,        
    citecolor=PineGreen,     
    filecolor=Plum,      	
    urlcolor=MidnightBlue,           
}



\def\lono{LiOsO$_3$/LiNbO$_3$}

\begin{document}
\title{Design of a Mott Multiferroic from a Non-Magnetic Polar Metal}

\author{Danilo Puggioni}
\affiliation{Department of Materials Science and Engineering, 
Northwestern University, 
IL 60208-3108, USA}
\author{Gianluca Giovannetti}
\affiliation{CNR-IOM-Democritos National Simulation Centre and International School
for Advanced Studies (SISSA), 
Trieste, Italy}
\author{Massimo Capone}
\affiliation{CNR-IOM-Democritos National Simulation Centre and International School
for Advanced Studies (SISSA), 
Trieste, Italy}
\author{James M.\ Rondinelli}
\affiliation{Department of Materials Science and Engineering, 
Northwestern University, 
IL 60208-3108, USA}

\begin{abstract}

We examine the electronic properties of newly discovered ``ferroelectric'' metal LiOsO$_3$ combining density-functional and dynamical mean-field theories. We show that the material is close to a Mott transition and  that electronic correlations can be tuned to engineer a Mott multiferroic state in  1/1 superlattice of LiOsO$_3$ and LiNbO$_3$. We use electronic structure calculations to predict that the (LiOsO$_3$)$_1$/(LiNbO$_3$)$_1$ superlattice is a type-I multiferroic material with a ferrolectric polarization of 41.2~$\mu$C cm$^{-2}$, Curie temperature of 927\,K, and N\'eel temperature of 671\,K. Our results support a route towards high-temperature multiferroics, \emph{i.e.}, driving non-magnetic \emph{polar metals} into correlated insulating magnetic states.
\end{abstract}

\pacs{%
75.85.+t, 
71.45.Gm, 
77.80.B$-$, 
71.20.$-$b,
}

\maketitle

\emph{Introduction}.---%
Multiferroics (MF) are a class of insulating materials where two (or more) primary ferroic order parameters, 
such as a ferroelectric polarization and long-range magnetic order, coexist. 
Technologically, they offer the 
possibility to control magnetic polarizations with  an electric field 
for reduced power consumption \cite{kimura, hur}.
Nonetheless, intrinsic room-temperature MF remain largely elusive.
This fact may be understood by examining 
the microscopic origins for the ferroic order which aids in classifying different phases: 
%
In Type-I MF, ferroelectricity and magnetism arise from different chemical species  
with ordering temperatures largely independent of one another and weak magnetoelectric (ME)
coupling \cite{wang}.
%
The ferroelectric ordering also typically appears at temperatures higher than 
the magnetic order, and the spontaneous polarization $P$ is 
large since it is driven by a second-order Jahn-Teller distortion, \emph{e.g.}, BiFeO$_3$ \cite{wang, pearson}. 
In Type-II MF, however, 
magnetic order induces ferroelectricity, 
which indicates a strong ME coupling between the two 
order parameters. 
Nonetheless, $P$ is usually much smaller, \emph{e.g.}, by a factor of 10$^2$  
as in $R$-Mn$_2$O$_5$ ($R$ being rare earth) \cite{cheong}.
In a few MFs with high-transition temperatures, \emph{i.e.}, BiFeO$_3$ \cite{Lee} and
Sr$_{1-x}$Ba$_x$MnO$_3$ \cite{Sakai,SBMOGG,Nourafkan}, magnetism is
caused by  Mott physics arising from strong correlations.
The interactions localize the spins at high temperature, paving the way for
magnetic ordering at room temperature. Materials where this
robust magnetism is coupled with ferroelectric distortions are 
ideal candidates for a room-temperature MFs.

Herein, we propose a design strategy for novel Mott MF phases. 
It relies on tuning the degree of correlation of the recently discovered class of 
materials referred to as `ferroelectric metals' with LiOsO$_3$ as the prototypical member \cite{Shi}.
This material is the first undisputed realization of the Anderson-Blount mechanism \cite{AndersonBlount}, and challenges the expectation 
that conduction electrons in metals would screen the electric field 
induced by polar displacements \cite{Shi, Yao, Sim}.
Despite robust metallicity, this material shares structural similarities with prototypical \emph{insulating} ferroelectric oxides, such as LiNbO$_3$ 
\cite{boysen, Megaw}: A  $R3c$ crystal structure with 
acentric cation displacements and distorted OsO$_6$ octahedra \cite{GGMC,Kim}
and comparable lattice parameters \cite{Shi, boysen}. 
While the \emph{polar displacements} in LiNbO$_3$ rely on cross-gap hybridization
between 
$p$ (O) and $d$ (Nb) states \cite{Inbar}, in LiOsO$_3$ they are weakly coupled to the states at the Fermi level ($E_F$), 
which makes possible the coexistence of an acentric structure and metallicity  \cite{PuggioniRondinelli,GGMC}.
In LiOsO$_3$ the empty $d$-manifold of LiNbO$_3$ is replaced by a 
non-magnetic $5d^3$ ground state with a half-filled $t_{2g}$ 
($d_{xy}$, $d_{xz}$, $d_{yz}$) configuration, which is responsible for the metallic response \cite{GGMC}. 
However, the strength of the electronic interactions is insufficient to drive a 
Mott transition in the correlated $t_{2g}$ manifold as revealed by low-temperature 
resistivity measurements; nonetheless, if it would be possible to 
enhance the electronic correlations in LiOsO$_3$ and achieve a metal-insulator 
transition, then a previously unidentified multiferroic material should result.
The concept is that if an insulating state can be obtained from 
a `ferroelectric metal' through enhanced correlations, it would then 
naturally lead to magnetic ordering of the localized electron spins, 
coexisting polar displacements, and 
potentially strong ME coupling.

In this work we explore the feasibility of this approach using a combination of first-principles 
density functional theory (DFT)  plus dynamical mean field theory (DMFT) calculations \cite{DMFT}. 
We first show that the electronic Coulomb interactions and Hund's coupling in LiOsO$_3$ 
make it an ideal candidate for realizing a Mott MF
due to the multi-orbital $t_{2g}$  physics. 
Next, we describe the design of a new multiferroic by control of the electronic 
structure through atomic scale engineering of a Mott metal-insulator
transition (MIT) in an ultrashort period (LiOsO$_3$)$_1$/(LiNbO$_3$)$_1$ superlattice.  
%
The insulating and magnetic state is driven by an enhancement of the 
electronic correlations in LiOsO$_3$ layers owing to the kinetic energy reduction 
of the $t_{2g}$ orbitals from the superlattice geometry. 
The ferroelectric properties mainly originate from cooperative Li and O displacements.
%
The multiferroic phase emerges across the MIT, exhibiting a net electric polarization (41.2\,$\mu$C\,cm$^{-2}$) and magnetization [0.9\,$\mu_B$ per formula unit (f.u.)], with calculated magnetic-ordering and ferroelectric temperatures of 671\,K and 927\,K, respectively.
Our results uncover a promising alternative route to discovery of room-temperature 
multiferroics: One could search for correlated \emph{polar metals near Mott transitions}  and drive the phases into insulating states, rather than the often-pursed approach of inducing polar displacements in robustly insulating magnets.
\emph{Calculation Details}.---%
We perform first-principles DFT calculations
within local-density approximation (LDA) $+$Hubbard $U$ method 
as implemented in the Vienna
{\it Ab initio} Simulation Package (VASP) 
\cite{Kresse} 
with the projector augmented wave (PAW) approach 
and a 600~eV plane wave cutoff 
with a $5\times7\times7$ $k$-point mesh.
%
We relax the volume and atomic positions (forces $<$0.1\,meV~\AA$^{-1}$) using Gaussian smearing (20\,meV width) for 
the  Brillouin zone (BZ) integrations.
We perform LDA+DMFT calculations including local Coulomb interactions
parameterized by the $U$ and the Hund's coupling $J_h$ starting from  
Wannier orbitals constructed from the LDA bands \cite{wannier90} using an energy range spanned by the full $d$ manifold.
The impurity model is solved using Exact Diagonalization (ED) 
with a parallel Arnoldi algorithm \cite{Capone,ARPACK}.

\emph{Correlations in LiOsO$_3$}.---%
We first examine the effect of the
interactions on the metallic state of LiOsO$_3$ and 
determine the critical values for a Mott transition $U_c$ in the paramagnetic and antiferromagnetic (AFM) phases using LDA+DMFT.
The criterion for a Mott-Hubbard transition is frequently associated with the 
ratio between the bandwidth ($W$) and the interaction
strength $U$,  so that the Mott transition occurs for $U_c$ of the order of $W$.
In a multiband Hubbard model with $M$ orbitals, $U_c$ is 
enhanced by orbital fluctuations, \emph{i.e.}, $U_c\sim\sqrt{M}W$, \cite{Gunnarsson} and it is influenced by the Hund's coupling $J_h$. 
Indeed, at half-filling, $U_c$ is \emph{reduced} by an enhancement of $J_h$ \cite{AGeorgesHund}. 

In the following, we show this is precisely the situation in LiOsO$_3$
\cite{GGMC}. 
Due to the energy separation between $t_{2g}$ and $e_g$ orbitals in the spectral density of state of LiOsO$_3$ around the Fermi level, we resort to using a model for the $t_{2g}$ levels only \cite{GGMC}. Owing to the symmetry breaking in bulk LiOsO$_3$, the orbitals in the $d$ manifold are also permitted to mix, which  lifts the degeneracy of $t_{2g}$ orbitals with two of states remaining degenerate.

\begin{figure}[t]%
\includegraphics[width=0.99\columnwidth,angle=-0]{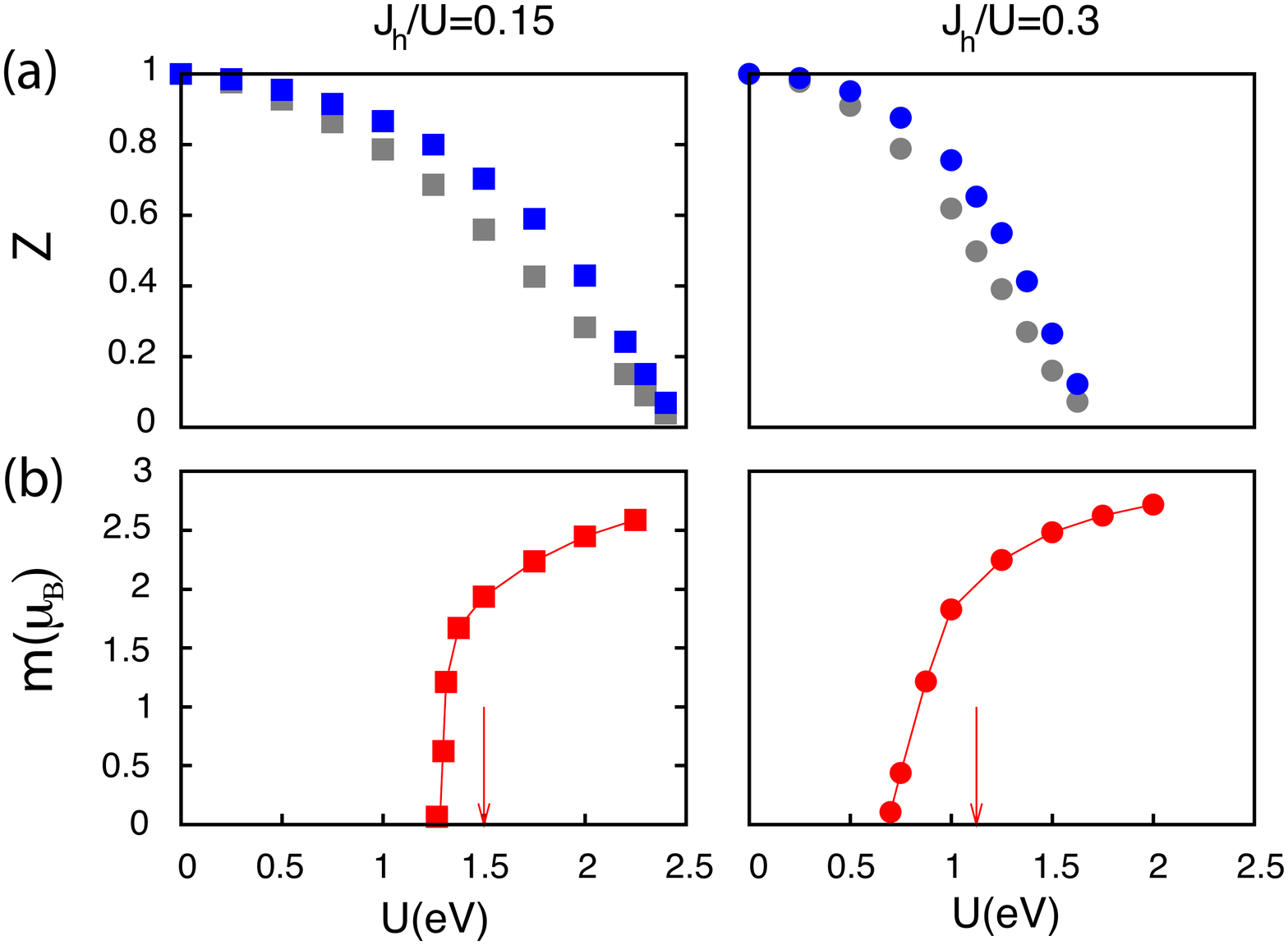}\vspace{-5pt}
\caption{(Color online) (a) Orbital resolved quasiparticle orbital weight Z (filled symbols) for paramagnetic  LiOsO$_3$ and (b) local magnetization $m$ ($\mu_B$) (obtained from a spin-polarized calculation) of the $t_{2g}$ orbitals in LiOsO$_3$ as function of $U$ for different ratios of $J_h/U$ within the LDA+DMFT  calculations.
Vertical arrows indicate the critical value of $U$ required to reach the insulating state in
the $G$-type AFM structure.}
\label{fig1}
\end{figure}

\autoref{fig1} shows the orbital resolved quasiparticle weight (Z) of the occupied orbitals as a 
function of $U$ for two different values of $J_h$ for paramagnetic LiOsO$_3$ in the experimental structure (see top panels). Z measures the metallic character of the system, and it evolves from Z=1 for a non-interacting metal to Z=0 for a Mott insulator.
Upon increasing the value of $J_h$, the critical value of $U$ required 
to reach the Mott state (Z=0) is shifted to larger values of $U$ \cite{AGeorgesHund}.

In the correlated regime, we anticipate electron localization will lead to long-range 
magnetic order of the localized spins. 
Spin-polarized LDA+DMFT calculations, initialized with a 
$G$-type AFM structure 
(every spin on an Os cation is antiparallel to all its neighbors), reveal that  
the local magnetic moment rapidly saturates to the atomic value $S=3/2$.
A finite magnetization also develops at intermediate $U$ in
the metallic state (\autoref{fig1}, lower panels). The MIT, marked by vertical arrows, occurs for a weaker coupling in the AFM than in the paramagnetic state.

\emph{Design of a Mott Multiferroic}.---%
The LDA+DMFT calculations reveal that a simultaneous Mott  
and  magnetic state could be engineered in LiOsO$_3$ 
by reducing the electronic kinetic energy. 
One avenue to control and decrease the kinetic energy relies on 
heterostructuring and interleaving two perovskites together 
to form a coherent superlattice, whereby an 
isostructural insulator would restrict the electron hopping due to the reduction 
in available channels \cite{gray,middey2012,middey2014}.
Such geometries  can be achieved in practice using 
oxide molecular-beam epitaxy or pulsed-laser deposition methods \cite{schlom, blok}.

\begin{figure}
\includegraphics[width=0.9\columnwidth]{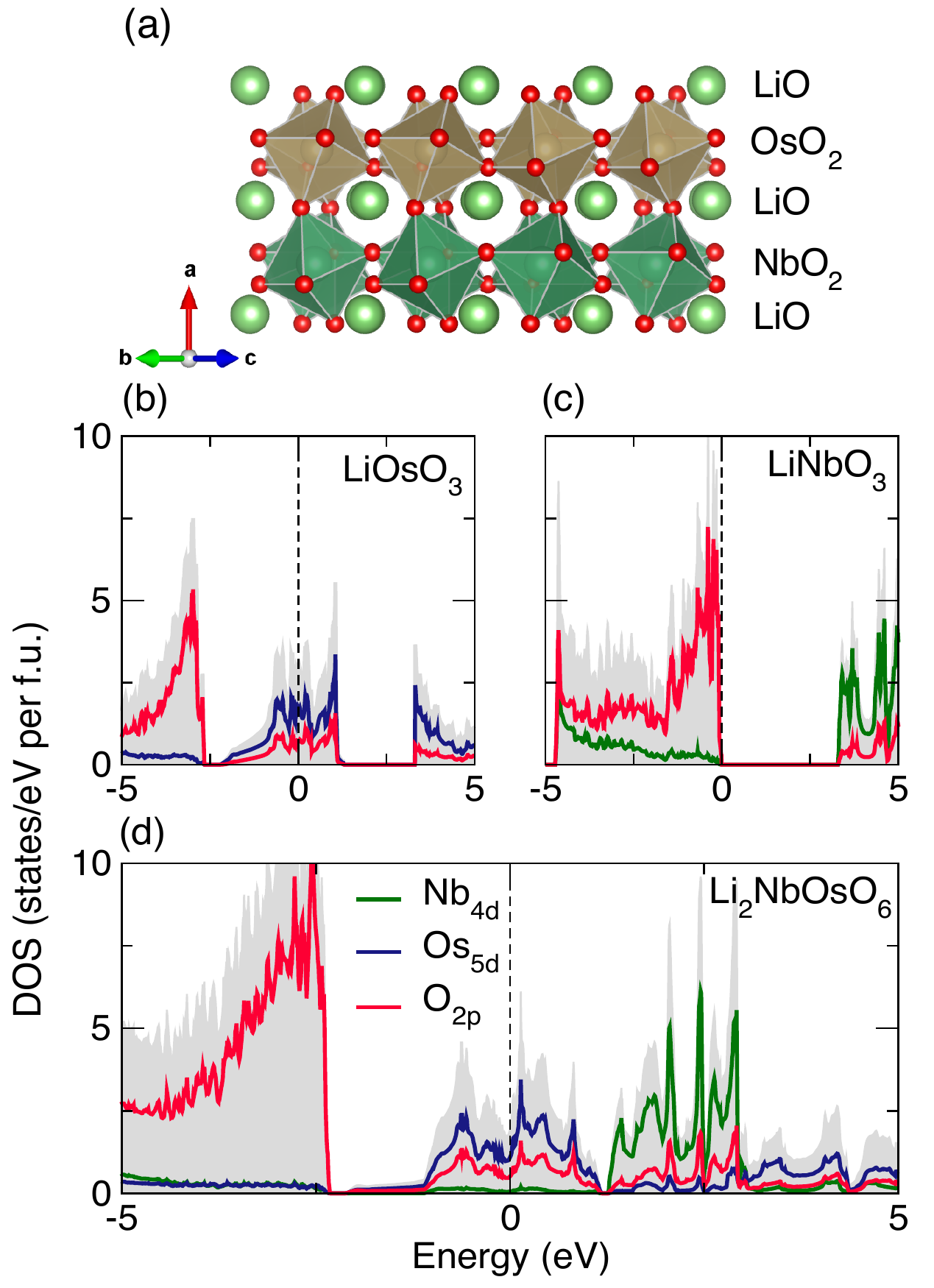}\vspace{-9pt}
\caption{(Color online) (a) The superlattice exhibits the $a^-b^-b^-$ tilt pattern. 
%
Atom- and orbital-resolved DOS for (b) LiOsO$_3$, (c) LiNbO$_3$ and (d) \lono\ at the DFT-LDA level. $E_F$ is given by the (broken) vertical line at 0\,eV.} 
\label{fig2}
\end{figure}

Owing to the chemical and structural compatibility of LiOsO$_3$ with
LiNbO$_3$, with a lattice mismatch of 3.2\%, we devise an 
ultrashort period perovskite superlattice of (LiOsO$_3$)$_1$/(LiNbO$_3$)$_1$ 
as illustrated in panel (a) of \autoref{fig2}.
The superlattice is constructed by beginning from the $R3c$ crystal structure of 
LiOsO$_3$ (LiNbO$_3$) and imposing a layered order along the [110] direction in 
the rhombohedral setting, which is equivalent to a 1/1 period
LiOsO$_3$/LiNbO$_3$ grown along the pseudocubic (pc) [001] direction
\cite{Note1}. 
The geometry in  \autoref{fig2} is also different from a superlattice constructed along the [101]$_\mathrm{pc}$ direction (\emph{c.f.}, Ref.~\onlinecite{xiang}), which is likely more challenging to realize experimentally. 
%
Following full relaxation of the superlattice, we find the cation order results in a symmetry
reduction to the polar space group $Pc$ with out-of-phase OsO$_6$ and NbO$_6$ octahedral rotations, \emph{i.e.}, the $a^-b^-b^-$ tilt pattern given in Glazer notation \cite{Glazer:1975}. 
The microscopic origin of polar displacements are described in detail below.

\emph{Electronic Properties}.---%
\autoref{fig2} shows the LDA electronic density of states (DOS) for
the \lono\ superlattice (d), compared with  
LiOsO$_3$ (b) and LiNbO$_3$ (c) using the LDA-optimized atomic structures.
The results for LiOsO$_3$ (\autoref{fig2}b) 
highlight the metallic character of the former, where the weight at
the Fermi level ($E_F$) mainly comes from Os $5d$ states which show strong
admixture from the O $2p$ states. 
In contrast, LiNbO$_3$ is a band insulator, with the O $2p$ states forming the 
valence band and Nb ${4d}$ states at the conduction band minimum, separated by 
a gap of 3.28\,eV (\autoref{fig2}c). 
In the superlattice, we find essentially no charge transfer between Os and Nb: 
Each component (LiOsO$_3$ and LiNbO$_3$) is isoelectronic to its bulk configuration;  
the DOS can be described as a direct superposition of the two components (\autoref{fig2}d).
The Os $5d$ states partially fill the gap in the electronic spectrum formed from the 
the two-dimensional NbO$_2$ planes. %
There is some spectral weight transfer in the vicinity 
of $E_F$ among the Os orbitals, which are sensitive to the electron correlation  
strength as shown in \autoref{fig1}.

\begin{figure}
\includegraphics[width=\columnwidth]{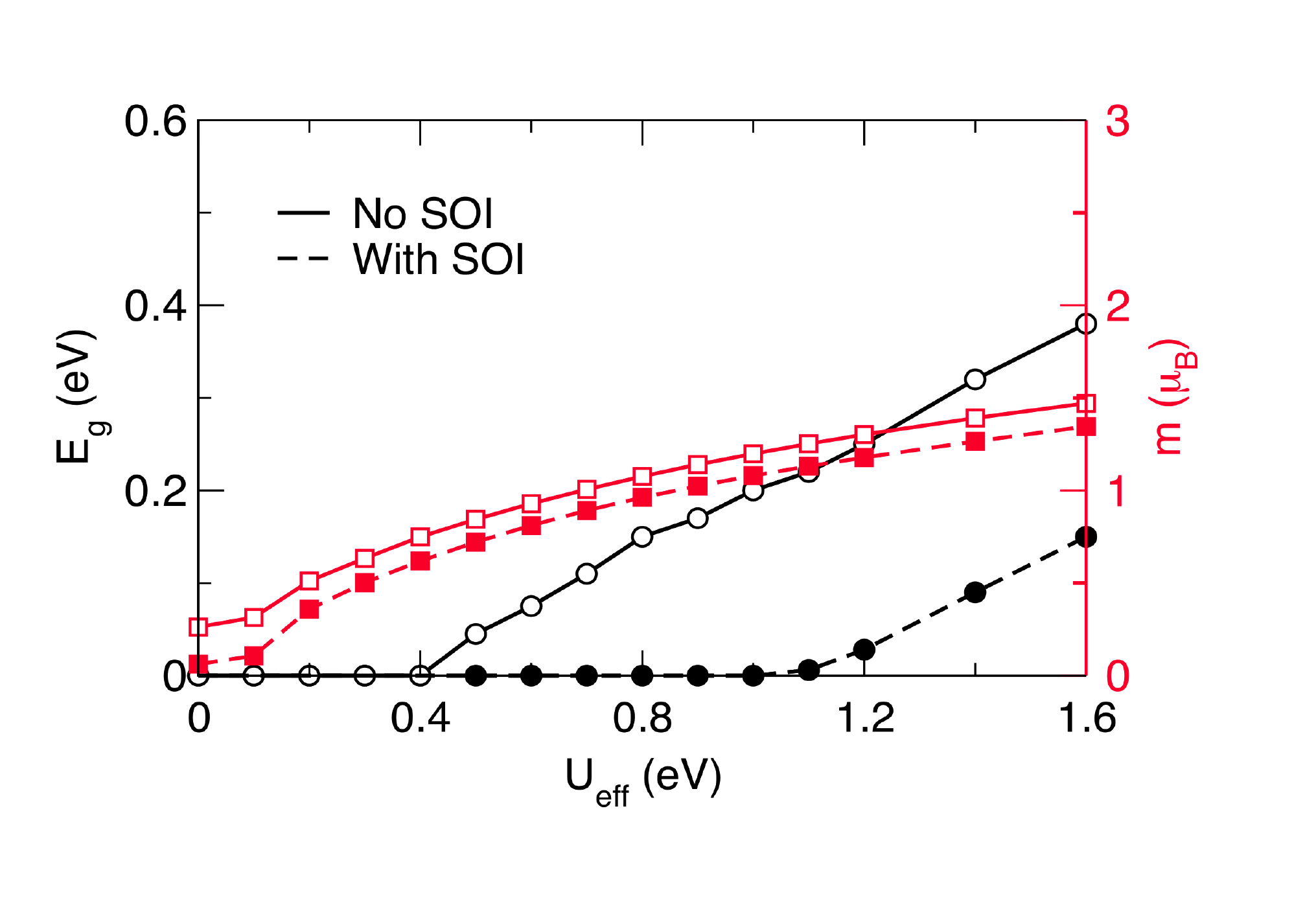}\vspace{-10pt}
\caption{(Color online) Band gap $E_g$ and averaged local magnetic moment for Os 
as a function of $U_\mathrm{eff}$ with and without spin-orbit interaction (SOI). }
\label{fig3}
\end{figure}

We now explore the effect of electronic correlations by means of LSDA$+U$ calculations at
different values of $U_\mathrm{eff}=U-J_h$.
An accurate value of the Hubbard $U$ is unknown for perovskite osmates, 
but it is expected to be comparable to that of NaOsO$_3$ \cite{Shi2} and 
double perovskite Sr$_2$CrOsO$_6$ \cite{Meetei} for which a correct description of the 
electronic properties are obtained with $U$ values of 1.0 and 2.0\,eV, respectively. 
%
Note that the differences from various implementations of the LDA$+U$ scheme for 
bulk LiOsO$_3$ were found to be minor \cite{GGMC}, and are anticipated to also be insignificant for the superlattice.

\autoref{fig3} shows the evolution in the band gap ($E_g$)  and magnetic moment of Os$^{3+}$ ions ($m$) as a function of the strength of $U_\mathrm{eff}$ for LSDA including spin-orbit interaction (SOI, broken lines). A gap opens at a critical 
 $U_\mathrm{eff} \sim 1$\,eV ($U_{c}$), 
signaling a MIT into a magnetic insulating ground state.
As expected the \lono\ superlattice becomes insulating for smaller
values of the interaction with respect to bulk LiOsO$_3$. 
The enhancement of electronic correlations is also found for small values of $U$. 
In fact, LiOsO$_3$ remains paramagnetic \cite{GGMC} while the
superlattice is weakly ferrimagnetic already below $U_{c}$, where it turns
into a  G-type AFM insulator at $U_\mathrm{eff} \sim0.5$\,eV.

The reduction in $U_{c}$ for the MIT in the superlattice 
can be understood by analyzing the
effect of the geometrical confinement on the $t_{2g}$ band dispersions. 
(For simplicity, we use the LDA electronic structures given in Ref.~\onlinecite{Note1}.)
While the bandwidth of the $d_{xy}$ orbitals is essentially the
same as for bulk LiOsO$_3$, the $d_{xz}$ and $d_{yz}$ bands in \lono\ are
significantly narrowed as a consequence of the reduced hopping along the
superlattice direction. This leads to a reduction of the kinetic
energy which enhances the electron-electron correlations, making the superlattice 
a Mott insulator at moderate interaction strengths.
We note that when SOI are excluded in the calculations (\autoref{fig3}, solid lines), 
the MIT occurs at a further reduced correlation strength ($U_{c}\sim0.5$\,eV), and the magnetic moment only slightly increases. Such behaviors are also observed in bulk LiOsO$_3$ \cite{GGMC}.


%
%

%

\begin{figure}
\includegraphics[width=0.65\columnwidth,angle=-90]{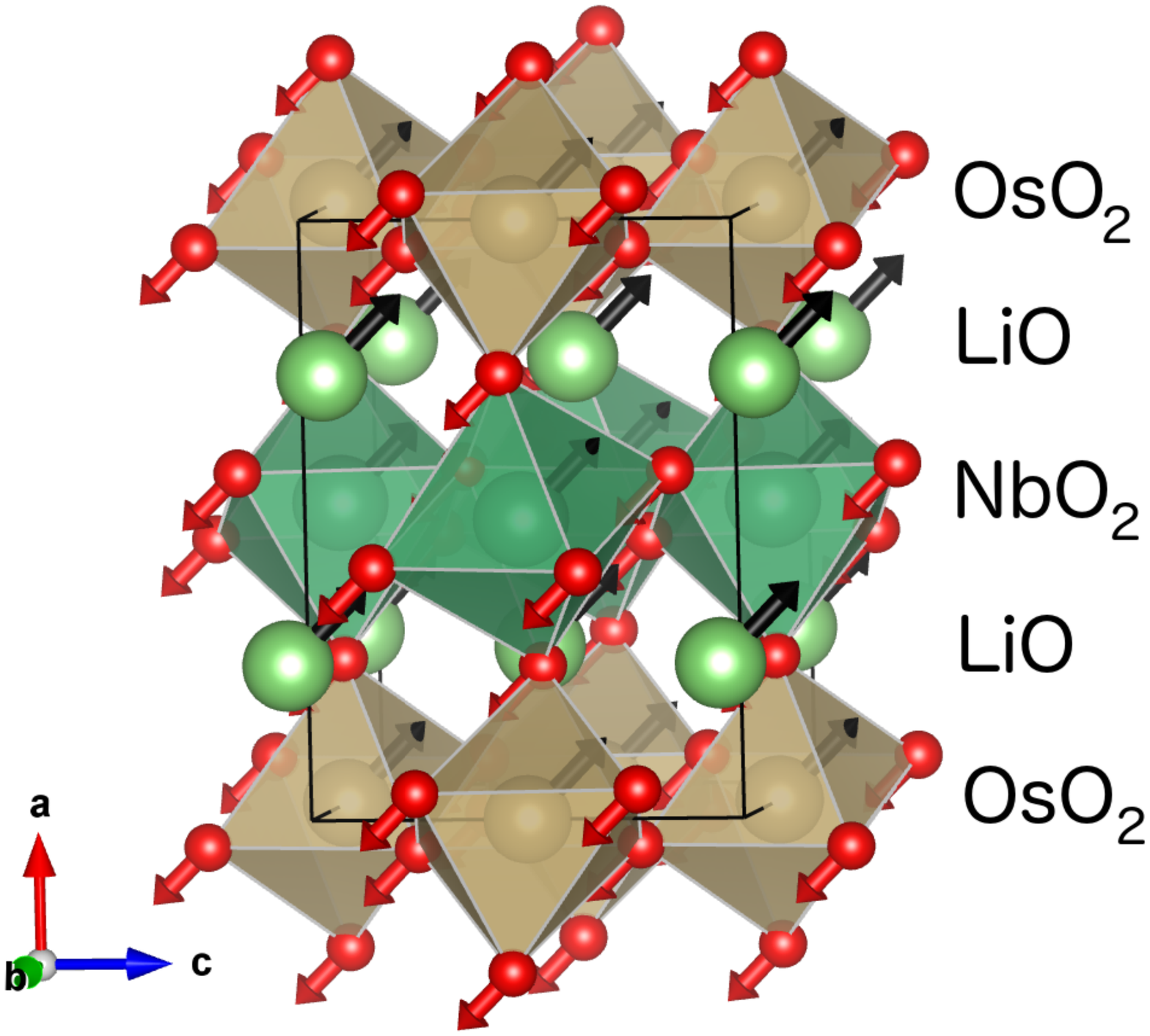}\vspace{-11pt}
\caption{(Color online) Illustration of the polar zone-center mode along the [101]-direction labeled by irrep $\Gamma_{2}^{-}$. Anti-polar displacements along the [010]-direction are omitted for clarity.}
\label{fig4}
\end{figure}

\emph{Ferroelectric Polarization}.---%
We now apply a group theoretical analysis \cite{Orobengoa:ks5225,Perez-Mato:sh5107} of the \lono\ structure to understand the inversion symmetry-breaking displacements that produce the $Pc$ ground state.
We use a fictitious $P2_1/c$ centrosymmetric phase (where polar displacements are switched off) as the reference phase from which the symmetry-adapted mode displacements are obtained as different irreducible representations (irreps) of the $P2_1/c$ space group  operators \cite{Note2}. 
We find the loss of inversion symmetry 
mainly derives from cooperative Li and O displacements in the (101) mirror plane of the 
$Pc$ phase.
Moreover, we find anti-polar displacements along the 
$b$-axis which result in no net polarization.  
All polar displacements are  described by a distortion vector that corresponds to the irrep  
$\Gamma_{2}^{-}$ along the [101]-direction of the $Pc$ structure (\autoref{fig4}). 
%
These displacements are consistent with the acentric Li and O ionic displacements 
identified to be responsible for lifting inversion symmetry in bulk LiOsO$_3$ \cite{Shi,Sim} 
and across the ferroelectric transition in LiNbO$_3$ \cite{benedek}.

We now compute the ferroelectric polarization in \lono\ using the Berry's phase 
approach \cite{BP} within LSDA+$U$ ($U_{c}=0.5$\,eV). The spontaneous electric 
polarization of the $Pc$ phase is 32.3\,$\mu$C\,cm$^{-2}$  and 25.5\,$\mu$C\,cm$^{-2}$ along the [100]-direction, \emph{i.e.,} along the pseudo-cubic [001] superlattice repeat 
direction and  [001]-directions, respectively. 
(Note that the [101]-direction in \lono\ corresponds to the polar [111]-direction in LiNbO$_3$.) 
Together this yields a net polarization along the [101]-direction of 41.2~$\mu$C\,cm$^{-2}$.
These values are also robust to SOI, with a change of less than 15\% to value of the total polarization. 
Following the recipe of Ref.~\onlinecite{jacek}, we use the energy difference between the high-symmetry ($P2_1/c$) and low-symmetry ($Pc$) to obtain a
ferroelectric Curie temperature of 927\,K for the superlattice.
This value is close to the extrapolated transition temperature for LiNbO$_3$ ($>$1,400\,K) \cite{Smolenskii}, and far exceeds that of bulk LiOsO$_3$ where inversion symmetry is lost near 140\,K \cite{Shi}.

\emph{Magnetic Ordering Temperature}.---%
Our DMFT calculations indicate that when the superlattice material enters in the
Mott state the magnetic moment is $\sim$3\,$\mu_B$, corresponding to a high-spin
$S=3/2$ state. 
We now estimate the N\'eel temperature for \lono\ by extracting the 
exchange interaction constants from spin-polarized DFT energies computed at $U_c$ without SOI 
following the approach in Ref.~\onlinecite{Lampis:2004}.
Assuming that the magnetism arises by ordering such
localized spins, we obtain intra- and inter-plane Os--Os exchange magnetic couplings
of -5.6\,meV and -0.2\,meV\, respectively, where a negative interaction indicate AFM exchange. 
From these values and without Anderson's renormalization \cite{Anderson}, we estimate a N\'eel temperature of 671\,K for the \lono\ superlattice, 
which makes the material a correlation-induced room-temperature multiferroic.

\emph{Conclusions}.---%
We used a LDA+DMFT approach to study the electronic
properties of the ``ferroelectric'' metal LiOsO$_3$. A detailed
understanding of the electronic structure of LiOsO$_3$ shows that a
reduction of the kinetic energy can drive the system into a Mott
insulating state. We use this concept to propose a strategy to design 
multiferroic materials by constructing a superlattice with the uncorrelated 
polar LiNbO$_3$ dielectric. 
On the basis of LSDA+U calculations we show that the ultra-short period 
\lono\ superlattice should be a type-I room-temperature Mott multiferroic with a large 
41.2~$\mu$C cm$^{-2}$ electric polarization. 

The large ferroelectric displacements from the LiNbO$_3$ layers 
facilitate the high ferroelectric ordering temperature in the \lono heterostructure as observed from the similarity in the Curie temperature of the superlattice with that of LiNbO$_3$. 
In this case \lono would behave as a paramagnetic 
Mott ferroelectric at high temperatures and transition into 
Mott multiferroic below the N\'eel temperature, 
which is predicted to be well-above room temperature.
Because the exchange interactions of Os are mediated by the coordinating O ligands, which are essential to and produce the ferroelectric distortion, a strong ME coupling is anticipated as in Sr$_{1-x}$Ba$_x$MnO$_3$ \cite{Sakai,SBMOGG,Nourafkan}.
We hope this work motivates the synthesis of new artificial multiferroics, and the adds to the growing discussion of new applications where noncentrosymmetric metals and ferroelectric materials may be united.

\begin{acknowledgments}
GG\ and MC\ acknowledge financial support by the European Research Council
under FP7/ERC Starting Independent Research Grant ``SUPERBAD" (Grant
Agreement No.\ 240524). 
DP and JMR acknowledge the ARO under Grant Nos.\  
W911NF-12-1-0133 and W911NF-15-1-0017 for financial support and the 
HPCMP of the DOD for computational resources. 
\end{acknowledgments}


\clearpage


\begin{widetext}
\begin{center}
{\bf\Large Design of a Mott Multiferroic from a Non-Magnetic Polar Metal \\
\bigskip
Supplementary Materials}
\end{center}
\end{widetext}


\begingroup
\begin{table*} 
\caption{\label{tab:lat_par_lioso}{Calculated crystallographic 
parameters for $R3c$ LiOsO$_3$ using LDA functional}. }
\begin{ruledtabular}
\centering
\begin{tabular}{lcccc}%
\multicolumn{5}{l}{\textbf{LiOsO$_3$} \hfill  $a=b=4.9399$~\AA, $c=13.3019$~\AA}  \\
\multicolumn{5}{l}{$R3c$ \hfill $\alpha=\beta=90^\circ$, $\gamma=120^\circ$}\\[0.4ex]
Atom	&	Wyck.\ Site	& $x$	& $y$	& $z$ 	\\
\hline
Li	&	$6a$	& 0		&	0	&	0.61117	\\
Os	&	$6b$	& 0		&	0	&	0.32084	\\
O	&	$18b$	&-0.00388	& 0.36259	& 0.06825\\[0.2em]
\end{tabular}
\end{ruledtabular}
\end{table*}
\endgroup

\begingroup
\begin{table*} 
\caption{\label{tab:lat_par_linbo}{Calculated crystallographic 
parameters for $R3c$ LiNbO$_3$ using LDA functional}.}
\begin{ruledtabular}
\centering
\begin{tabular}{lcccc}%
\multicolumn{5}{l}{\textbf{LiNbO$_3$} \hfill  $a=b=5.1052$~\AA, $c=13.7472$~\AA}  \\
\multicolumn{5}{l}{$R3c$ \hfill $\alpha=\beta=90^\circ$, $\gamma=120^\circ$}\\[0.4ex]
Atom	&	Wyck.\ Site	& $x$	& $y$	& $z$ 	\\
\hline
Li	&	$6a$	& 0		&	0	&	0.61517	\\
Nb	&	$6b$	& 0		&	0	&	0.33169	\\
O	&	$18b$	&-0.01367	& 0.35987	& 0.06330\\[0.2em]
\end{tabular}
\end{ruledtabular}
\end{table*}
\endgroup

\begingroup
\begin{table*} 
\caption{\label{tab:lat_par_liosnbo}{Calculated crystallographic 
parameters for $Pc$ Li$_2$NbOsO$_6$ using the LDA functional}.}
\begin{ruledtabular}
\centering
\begin{tabular}{lcccc}%
\multicolumn{5}{l}{\textbf{LiOsO$_3$/LiNbO$_3$} \hfill  $a=7.35270$~\AA,  $b=5.02739$~\AA, $c=5.36541$~\AA}  \\
\multicolumn{5}{l}{$Pc$ \hfill $\alpha=\gamma=90^\circ$, $\beta=95.109^\circ$}\\[0.4ex]
Atom	&	Wyck.\ Site	& $x$	& $y$	& $z$ 	\\
\hline
Li1   & $2a$ &0.27896&  0.25751& 0.45909  \\
Li2    &$2a$ &0.78342  &0.74972 &-0.07013 \\
Nb    &$2a$ &-0.00633 &0.25287 &0.00169  \\
Os    &$2a$ &0.49713  &0.75123 &0.53150  \\
O1    &$2a$ &0.23965  &0.12376 &0.02510  \\
O2     &$2a$ &0.73578  &0.60514 &0.54599  \\
O3      &$2a$ &-0.09599 &0.05685 &0.71217  \\
O4      &$2a$ &0.41768  &0.54538 &0.23192  \\
O5      &$2a$ &0.04578  &0.58866 &0.85929  \\
O6      &$2a$ &0.54412  &0.04168 &0.32305  \\[0.2em]
\end{tabular}
\end{ruledtabular}
\end{table*}
\endgroup

\begingroup
\begin{table*} 
\caption{\label{tab:lat_par_liosnbo_nomono}{Calculated crystallographic 
parameters for $Pc$  Li$_2$NbOsO$_6$ without monoclinic angle using the LDA functional}.}
\begin{ruledtabular}
\centering
\begin{tabular}{lcccc}%
\multicolumn{5}{l}{\textbf{LiOsO$_3$/LiNbO$_3$} \hfill  $a=7.35884$~\AA,  $b=5.03159$~\AA, $c=5.36989$~\AA}  \\
\multicolumn{5}{l}{$Pc$ \hfill $\alpha=\beta=\gamma=90^\circ$}\\[0.4ex]
Atom	&	Wyck.\ Site	& $x$	& $y$	& $z$ 	\\
\hline
Li1 &   $2 a$ &0.27310 & 0.26765 &0.47516  \\
Li2  &  $2 a$ &0.78081  &0.75357 &-0.05429 \\
Nb   & $2 a$ &-0.00421 &0.25110 &0.00188  \\
Os    &$2 a$ &0.49815  &0.75234 &0.52637  \\
O1     & $2 a$ &0.24071 & 0.12246 &0.01238  \\
O2      &$2 a$ &0.73609  &0.60651 &0.54411  \\
O3      &$2 a$ &-0.09326 &0.05556 &0.71992  \\
O4      &$2 a$ &0.41773  &0.53696 &0.24015  \\
O5      &$2 a$ &0.04603  &0.58671 &0.85024  \\
O6      &$2 a$ &0.54504  &0.03169 &0.30376  \\[0.2em]
\end{tabular}
\end{ruledtabular}
\end{table*}
\endgroup

\begingroup
\begin{table*} 
\caption{\label{tab:lat_par_liosnbo_ldau}{Calculated crystallographic 
parameters for $Pc$ Li$_2$NbOsO$_6$  without monoclinic angle using the LSDA+$U$ with $U$=$U_{c}$=0.5~eV}.}
\begin{ruledtabular}
\centering
\begin{tabular}{lcccc}%
\multicolumn{5}{l}{\textbf{LiOsO$_3$/LiNbO$_3$} \hfill  $a=7.35714$~\AA,  $b=5.03042$~\AA, $c=5.36865$~\AA}  \\
\multicolumn{5}{l}{$Pc$ \hfill $\alpha=\beta=\gamma=90^\circ$}\\[0.4ex]
Atom	&	Wyck.\ Site	& $x$	& $y$	& $z$ 	\\
\hline
Li1  &  $2 a$ &0.27431 & 0.26460 &0.47149  \\
Li2   & $2 a$ &0.77680  &0.75646 &-0.05271 \\
Nb    &$2 a$ &-0.00460 &0.25150 &0.00212  \\
Os    &$2 a$ &0.49708  &0.75063 &0.52604  \\
O1     & $2 a$ &0.24114 & 0.12241 &0.01487  \\
O2      &$2 a$& 0.73596  &0.60719 &0.54175  \\
O3      &$2 a$ &-0.09280 &0.05536 &0.72002  \\
O4      &$2 a$ &0.41875  &0.54274 &0.23604  \\
O5      &$2 a$ &0.04702 & 0.58598 &0.84961  \\
O6      &$2 a$& 0.54653  &0.03942 &0.31046  \\[0.2em]
\end{tabular}
\end{ruledtabular}
\end{table*}
\endgroup


\begin{figure*} 
\includegraphics[width=0.75\textwidth]{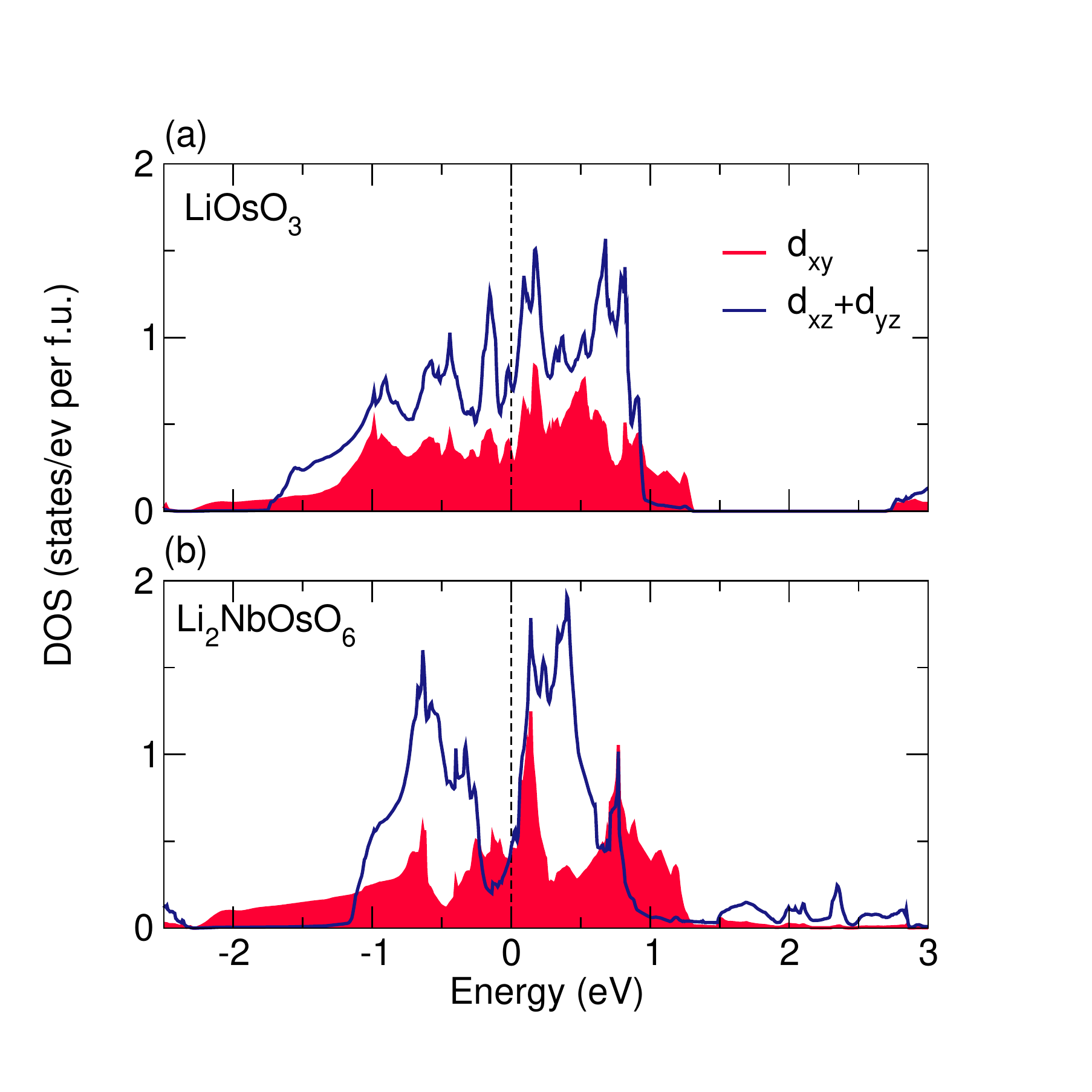}
\caption{(Color online) Resolved 5$d$ t$_{2g}$ states of Os for (a) LiOsO$_3$ and (b) Li$_2$NbOsO$_6$ within LDA. The bandwidth of the $d_{xy}$ orbitals  for bulk LiOsO$_3$ and Li$_2$NbOsO$_6$  are the same while in the superlattice the $d_{xz}$ and $d_{yz}$ orbitals have a reduced bandwidth. 
}
\label{figs1}
\end{figure*}

\begin{figure*} 
\includegraphics[width=0.75\textwidth]{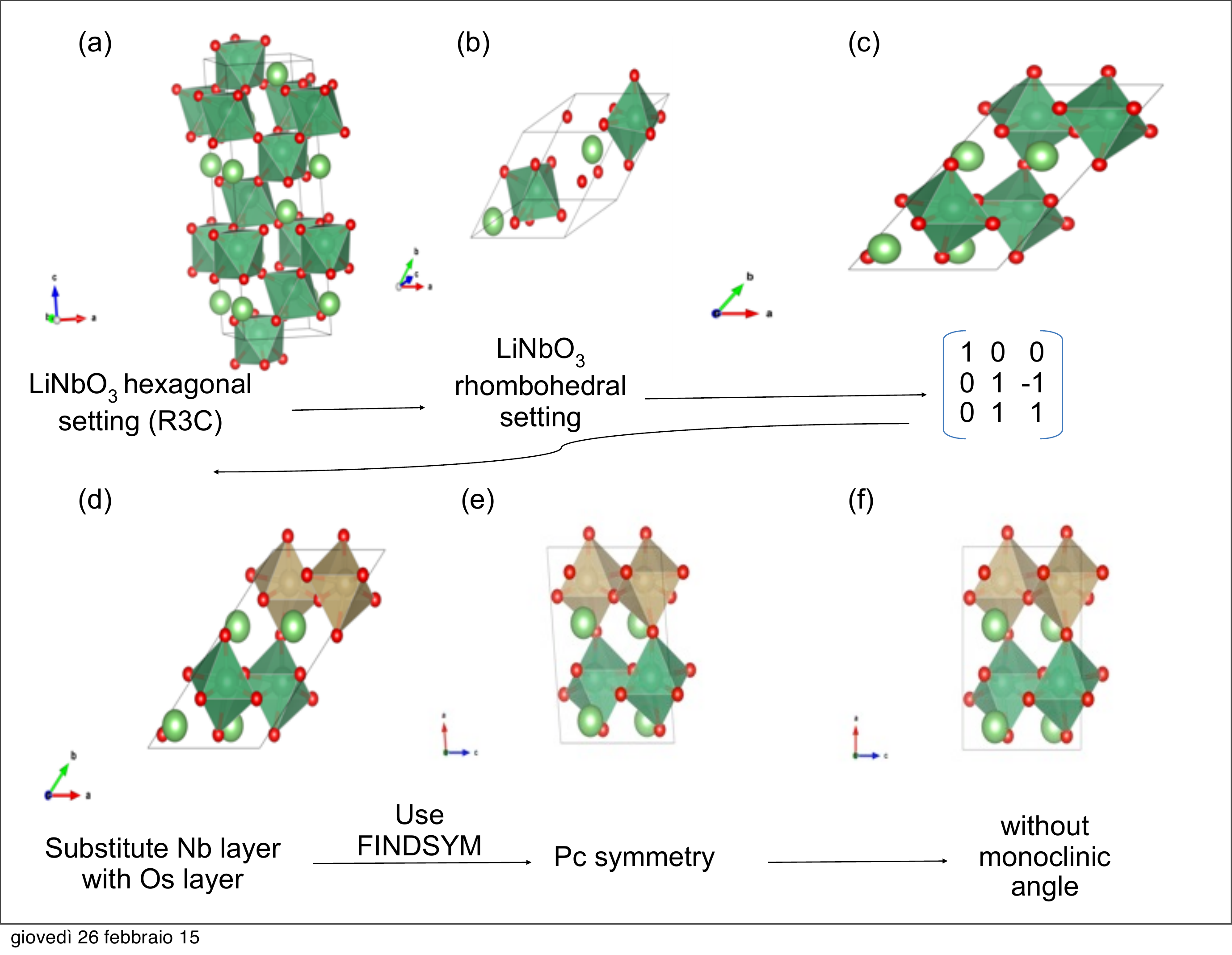} 
\caption{(Color online) Superlattice construction: (a) Starting from the $R3c$ crystal structure of LiNbO$_3$ in the hexagonal setting  with 30 atoms, (b) we transform to the rhombohedral setting with 10 atoms. (c) Next  we double the cell with the transformation matrix and (d) substitute a NbO$_2$ layer with OsO$_2$ layer. (e) We identify the new space group $Pc$ with FINDSYM \cite{findsym} and finally (f) we set the monoclinic angle to 90$^\circ$.
}
\label{figs2}
\end{figure*}

\end{document}